\documentclass[runningheads]{llncs}
\usepackage[table]{xcolor}
\usepackage[margin=1in]{geometry}
\usepackage{tabularx}
\usepackage{enumitem}
\usepackage{algorithm}
\usepackage{algorithmic}
\usepackage{dirtree}
\usepackage{multicol}

\setlist{nolistsep}

\definecolor{blue}{HTML}{008ED7}
\definecolor{mygray}{gray}{0.75}
\definecolor{lightBlue}{HTML}{e5f7ff}

\usepackage{graphicx}
\usepackage{cite}

\begin{document}

\title{PoCaP Corpus: A Multimodal Dataset for Smart Operating Room Speech Assistant using Interventional Radiology Workflow Analysis\thanks{Seung Hee Yang is the corresponding author.}}

\titlerunning{PoCaP Corpus: A Multimodal Dataset for Smart Operating Room Speech Assistant}

\author{ Kubilay Can Demir\inst{1}\orcidID{0000-0002-8084-753X } \and
Matthias May \inst{2} \and Axel Schmid \inst{2} \and Michael Uder\inst{2} \and
Katharina Breininger \orcidID{0000-0001-7600-5869} \inst{3} \and Tobias Weise \inst{1,4} \orcidID{0000-0001-6225-1348} \and Andreas Maier\inst{4}\orcidID{0000-0002-9550-5284} \and
Seung Hee Yang \inst{1,*}\orcidID{0000-0002-5004-1615}}

\institute{ Speech \& Language Processing Lab., Friedrich-Alexander-Universität Erlangen-Nürnberg \\ Erlangen, Germany \\ \email{kubilay.c.demir@fau.de} \and Radiologisches Institut Universitätsklinikum Erlangen,  Erlangen, Germany \and Artificial Intelligence in Medical Imaging Lab., Friedrich-Alexander-Universität Erlangen-Nürnberg \and Pattern Recognition Lab, Friedrich-Alexander-Universität Erlangen-Nürnberg \\  Erlangen, Germany}

\authorrunning{Demir et al.}

\maketitle

\begin{abstract}
This paper presents a new multimodal interventional radiology dataset, called PoCaP (Port Catheter Placement) Corpus. This corpus consists of speech and audio signals in German, X-ray images, and system commands collected from 31 PoCaP interventions by six surgeons with average duration of $81.4 \pm 41.0$ minutes. The corpus aims to provide a resource for developing a smart speech assistant in operating rooms. In particular, it may be used to develop a speech-controlled system that enables surgeons to control the operation parameters such as C-arm movements and table positions. In order to record the dataset, we acquired consent by the institutional review board and workers' council in the University Hospital Erlangen and by the patients for data privacy. We describe the recording set-up, data structure, workflow and preprocessing steps, and report the first PoCaP Corpus speech recognition analysis results with 11.52\% word error rate using pretrained models. The findings suggest that the data has the potential to build a robust command recognition system and will allow the development of a novel intervention support systems using speech and image processing in the medical domain. 


\end{abstract}

\section{Introduction}
Modern operating rooms (OR) are adapting advancing technologies rapidly and becoming more digitalized environments \cite{maier2017surgical}. Necessary medical devices are collecting and visualizing large amounts of data required for an improved execution of an operation. This allows physicians to do more intricate and successful procedures, improving patient safety and OR efficiency. However, the quantity of data and the complexity of an operation cannot be expanded indefinitely. Therefore, intelligent systems which can follow the execution of an operation and assist physicians are proposed \cite{herfarth2003lean}. These systems can process available data in the OR and present it in the correct time and format, follow the operation semantically, and take over some routine tasks.

A major approach in the creation of an intelligent workflow assistance system is surgical workflow analysis, which is often done by hierarchically decomposing an operation into smaller meaningful activities on different semantic levels \cite{lalys2014surgical}. In this approach, an operation is typically defined by \textit{phases}, \textit{steps} or \textit{actions}. Phases are highest-level actions such as preparation, cutting, or sterilization. Steps are necessary activities to perform phases, such as table positioning, instrument setting, or putting covers. Actions are generally the lowest-level activities in an analysis. They include basic activities such as grabbing an instrument or turning on a device. 

The available corpora for surgical workflow analysis vary considerably in size, quality, coverage, and depth of linguistic and structural characteristics. However, the vast majority of corpora only contains endoscopic video data, while some include additional instrument usage information. Other possible data modalities in OR such as speech are mostly under-investigated. We claim that a corpus for surgical workflow analysis can benefit from multimodal data, including not only images and videos, but also speech uttered by the surgeons in their language. This will allow development of a smart OR speech assistant, which is able to understand the different phases of a surgery and identify surgeons’ command words during operations.

To this end, we construct a new kind of multimodal German dataset in the domain of interventional radiology workflow. We describe the design, collection, and development of our 31 PoCaP (Port-catheter Placement) Corpus. It consists of X-ray images, RGB ambient videos, speech uttered by the surgeons, and system commands given by medical personnel to operate devices in the OR. The dataset is annotated with different levels of structural and descriptive meta-information, as well as linguistic information. All speech samples are automatically transcribed using an Automatic Speech Recognition (ASR) engine, and surgical phases are manually annotated with the help of surgeons in the Radiology Department at the University Hospital of Erlangen in Germany. Unfortunately, this dataset cannot be made publicly available to ensure patients' and personal's data privacy. 

The following section of this paper summarizes the related works on biomedical corpus collection and processing. Section 3 and 4 describe our PoCaP data and preprocessing results, respectively. In Section 5, we discuss surgery-specific and surgery-independent challenges regarding annotation and data processing. Finally, we end with the potential use of the corpus, future research direction, and concluding remarks. 
 
\section{Related Works}
Similar to other medical fields, publicly available data for surgical phase recognition are not abundant. Firstly, collecting and annotating data from OR is a costly procedure. Acquiring any sort of data requires the implementation of new hardware or software in the OR and maintenance during the whole procedure. Annotation is a time-intensive step in the data collection and needs to be performed or validated by medical experts. Secondly, local legal requirements have to be followed in order to ensure the security of the medical data. This factor may also limit the public sharing of collected datasets.

\textit{Cholec80} is a endoscopic video dataset of 80 laparoscopic cholecystectomy surgeries, including phase annotations for seven distinctive phases and seven tools \cite{twinanda_endonet_2017}. \textit{m2cai16-workflow} dataset contains 41 endoscopic videos of laparoscopic cholecystectomy and phase annotations for eight phases \cite{stauder2016tum, twinanda_endonet_2017}. \textit{HeiChole} is another endoscopic video dataset of laparoscopic cholecystectomy containing 33 surgeries with phase, action and tool annotations \cite{wagner_comparative_2021}. In HeiChole, phase annotations are provided similar to Cholec80. In \cite{bar2020impact}, authors created a private dataset of 1243 endoscopic videos for laparoscopic cholecystectomy from multiple medical centers. \textit{CATARACTS} is a microscopic video dataset of 50 cataracts surgery, which contains annotations consisting of 14 phases \cite{zisimopoulos_deepphase_2018}. Another dataset of cataract surgery is \textit{Cataract-101}, which has 101 operations and annotations with ten phases \cite{schoeffmann_cataract101}. \textit{Kitaguchi et al.}  collected 300 laparoscopic colorectal surgery videos from multiple institutions and annotated them with nine phases \cite{kitaguchi2020automated}. \textit{Bypass40} is an private dataset of 40 endoscopic videos of gastric bypass surgical procedures annotated with both phases and steps \cite{ramesh2021multi}.

\section{Data Collection Procedure and Setup}

\begin{figure}
    \centering
    \includegraphics[width = \textwidth]{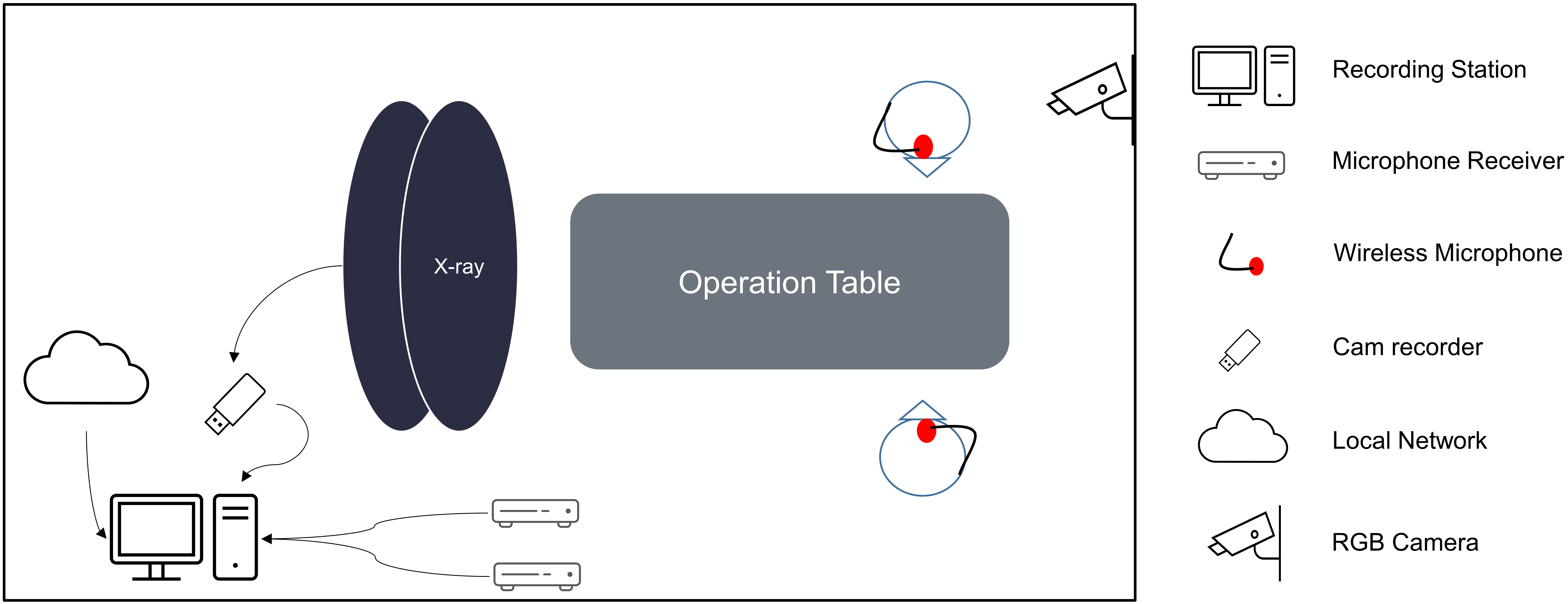}
    \caption{The data recording setup of our corpus. During an intervention, X-ray images are recorded with a C-Arm machine and displayed on a screen in the OR. X-ray images were captured from this display. The surgeon and the assistant were equipped with personal microphones to record their speech signals. System commands were recorded from a local network utilizing a monitoring software. An RGB camera with an embedded microphone recorded the interventions. Audio data from this camera were added to our corpus and recorded videos are used solely for aiding with the annotation process.}
    \label{fig:recording}
\end{figure}

As described in the previous section, datasets on the workflow analysis are mostly concentrated on video signals, either endoscopic or microscopic. Some datasets include additional tool annotations. However, other possible data sources are mostly under-investigated. In this paper, we propose a multimodal dataset consisting of three-channel speech and audio signals, X-ray images, and system commands collected during PoCaP interventions in the Radiology Department of University Hospital Erlangen, Germany. Before data collection, we obtained approvals from the institutional review board and workers' council. Additionally, we asked every patient for their consent in a written form. Operations are performed by six different surgeons with different levels of expertise. The average duration of an intervention is $81.4 \pm 41.0$ minutes. Additionally, we have captured ambient videos to help the annotation procedure, see fig \ref{fig:recording}. Our dataset includes 31 operations. We defined 31 surgical steps and eight surgical phases for the PoCaP intervention, to use in surgical phase and step recognition tasks, which are shown in Table \ref{table:phases}.

\begin{table}
\caption{Definitions of surgical phases and steps of port-catheter placement surgery in the PoCaP Corpus.}
\label{table:phases}
\fontfamily{cmss}\selectfont
\begin{tabularx}{\textwidth}[t]{XX}
Phases & Steps \\
\hline

1. Preparation & 
\begin{minipage}[t]{\linewidth}%
\begin{itemize}
\item[1.1] Patient positioning on the table
\item[1.2] Table moves up
\item[1.3] Radiologist in sterile
\item[1.5] Preparation of sterile material 
\item[1.6] Patient in sterile
\vspace{1mm}
\end{itemize} 
\end{minipage}\\
\hline

2. Puncture &
\begin{minipage}[t]{\linewidth}%
\begin{itemize}
\item[2.1] Local anesthesia 
\item[2.2] Ultrasound guided puncture
\end{itemize} 
\end{minipage}\\
\hline

3. Positioning of the Guide Wire &
\begin{minipage}[t]{\linewidth}%
\begin{itemize}
\item[3.1] C-Arm moves in
\item[3.2] Fluoroscopy on the subclavian area
\item[3.3] Fluoroscopy on the vena cava inferior (VCI) area
\item[3.4] C-Arm moves out
\end{itemize}
\end{minipage}\\
\hline

4. Pouch Preparation and Catheter Placement &
\begin{minipage}[t]{\linewidth}%
\begin{itemize}
\item[4.1] Local anaesthesia
\item[4.2] Incision
\item[4.3] Pouch preparation
\item[4.4] Sheath
\end{itemize}
\end{minipage}\\
\hline

5. Catheter Positioning &
\begin{minipage}[t]{\linewidth}%
\begin{itemize}
\item[5.1] C-Arm moves in
\item[5.2] Fluoroscopy on the VCI area
\item[5.3] Positioning of the catheter
\end{itemize}
\end{minipage}\\
\hline

6. Catheter Adjustment &
\begin{minipage}[t]{\linewidth}%
\begin{itemize}
\item[6.1] Shortening of the catheter
\item[6.2] C-Arm moves out
\item[6.3] Connection of the catheter to the port capsule
\item[6.4] Positioning of the port capsule in the pouch
\item[6.5] Surgical suture
\item[6.6] Puncture of the port capsule
\end{itemize}
\end{minipage}\\
\hline

7. Catheter Control &
\begin{minipage}[t]{\linewidth}%
\begin{itemize}
\item[7.1] C-Arm moves in
\item[7.2] Digital subtraction angiography of the chest
\item[7.3] C-Arm moves out to parking position
\end{itemize}
\end{minipage}\\
\hline

8. Closing &
\begin{minipage}[t]{\linewidth}%
\begin{itemize}
\item[8.1] Sterile patch
\item[8.2] Table moves down
\end{itemize}
\end{minipage}\\

\end{tabularx}
\end{table}

\subsection{Port-Catheter Placement}
Port-catheter placement is a frequently executed intervention in the radiology department. It is often applied to patients who require frequent infusions, e.g. during chemotherapy. A port-catheter is a device consisting of a port, an access point to use for infusions during treatment, and a thin flexible tube called a catheter. During an intervention, the port is placed approximately two centimetres under the skin on either side of the chest and the catheter is connected to a large vein emptying into the heart. The placement of a port makes it possible to avoid repeated injuries to small vessels and chemotherapy-related inflammation of peripheral veins during chemotherapy \cite{gonda2011principles}. 

\subsection{Data Collection Setup}
A PoCaP intervention is commonly performed by a surgeon and an assistant. While the surgeon is in the sterile area, the assistant controls the table, lights, C-Arm X-ray machine, or other medical equipment in accordance with commands given by the surgeon. Therefore, it is meaningful to capture all verbal conversations between the surgeon and the assistant, in order to capture rich contextual information for the surgical workflow analysis. Additionally, speech is the sole data source that appears in every type of surgery, and it is therefore important to analyze for the design of scalable systems. In order to record these speech signals, the surgeon and the assistant were equipped with personal microphones during surgeries. For this task, we used two sets of Sennheiser XSW 2 ME3-E wireless headsets. With this configuration, we were able to obtain two-channel high-quality speech signals uttered by the surgeon and the assistant individually. In some cases, a second surgeon or assistant was necessary in the OR, however, that personnel didn't wear a microphone. The necessity is generally due to complications during the operation.

\begin{figure}
    \centering
    \includegraphics[scale = 0.3]{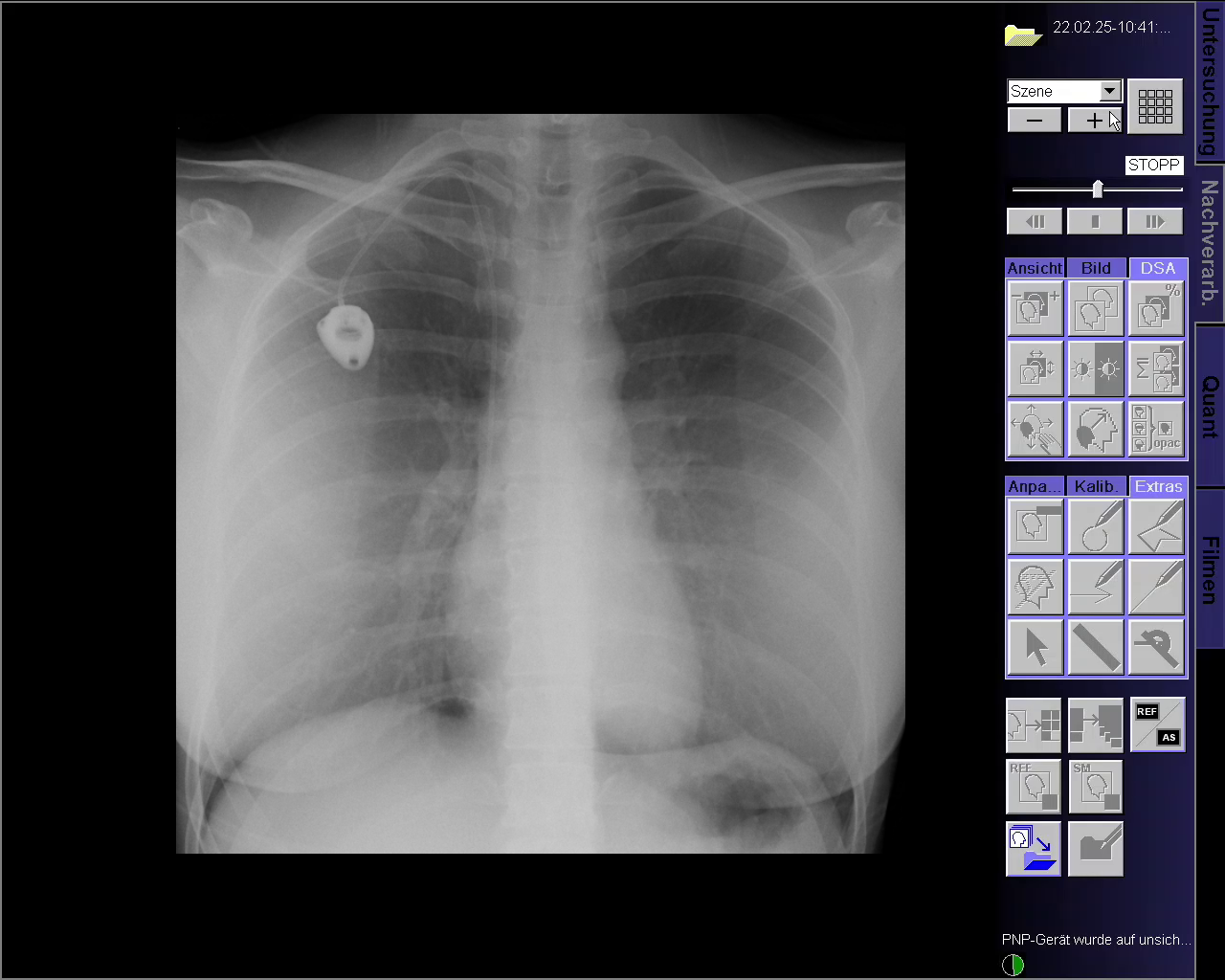}
    \caption{An illustration of the screen capture with an X-ray image from a port-catheter placement intervention captured with OBS. For data privacy purposes, the X-ray image is replaced by an open-source image from \cite{xray}.}
    \label{fig:image}
\end{figure}

X-ray images are another highly useful data source, especially for a PoCaP intervention. For example, the subclavian area and VCI area are monitored before catheter placement, using fluoroscopy during the intervention. Furthermore, the catheter is guided into the patients' body and controlled similarly by X-ray images. These activities are represented in phases 3,5 and 7 in Table \ref{table:phases}. For those reasons, it is a very rich complementary data source for the surgical workflow analysis in this case. In the OR, X-ray images are shown on a widescreen located on the other side of the surgical table for immediate assessment. To record X-ray data, we utilized the input signal going into this display, see Fig. \ref{fig:image}. We duplicated the input signal and recorded it with the Open Broadcaster Software (OBS) \cite{Obs2022}. We also utilized OBS to concurrently capture audio and speech input from the configuration described in the previous paragraph.

Modern ORs have many digital sensors providing information about the status and activities of medical devices. We included such data from the C-Arm machine in our corpus as system commands. System commands refer to status and activity logs and they are created when the device is utilized, which is typically done by the assistant during a PoCaP intervention. In our case, the C-Arm machine is connected to a local network and can be monitored by a local PC and a monitoring software called \textit{ACSOS}. ACSOS generates a \textit{ticket} when a function of a medical device in the OR is used and logs all related variables. It can display a total of 393 different variables in total. An example of a ticket is shown in Fig.~\ref{fig:axcs}. This information can be regarded as similar to tool usage information and as a relevant source for the workflow analysis. We manually copied this data from the local computer for each operation.

\begin{figure}
\centering
\begin{tabular}{ |p{4cm}|p{3cm}|p{2cm}|p{2cm}|}
\hline
\multicolumn{4}{|c|}{Ticket: C-Arm Machine Status} \\
\hline
Variable& Min - Max Range & Dimension & Value \\
\hline
Longitudinal position & [-5000, 5000] & mm & 0 \\
Height from floor & [-1000, 5000] & mm & 1100 \\
Physical rotation angle & [-3600, 3600] & 0.1 Degree & 510 \\
Acquisition and fluro mode & [1, 20] & none & 5 \\
X-ray pulse rate & [0, 32767] & 0.01 Frame / second & 7100 \\
\hline
\end{tabular}
     \caption{When the C-Arm machine is utilized, a respective ticket containing all necessary variables is created. The exemplary ticket variables in this table show the current position of the C-arm machine and mode settings with arbitrary values.}
     \label{fig:axcs}
 \end{figure}

In order to ease the annotation procedure, we recorded ambient videos with a single GoPro Hero 8 camera. Depending on the patients' choice, intervention can be done on the left or right side of their chest. We positioned the camera on the left or right side of the OR to cover the operation table and the sterile area. We utilized ambient videos solely to annotate our corpus accurately. However, these ambient videos include also speech and audio signals recorded with an embedded microphone. We integrated these signals as a third channel to our dataset. When compared to personal microphones, this data channel is noisy and reverberant. 

\section{Dataset}
\subsection{Data Structure}
\begin{figure}
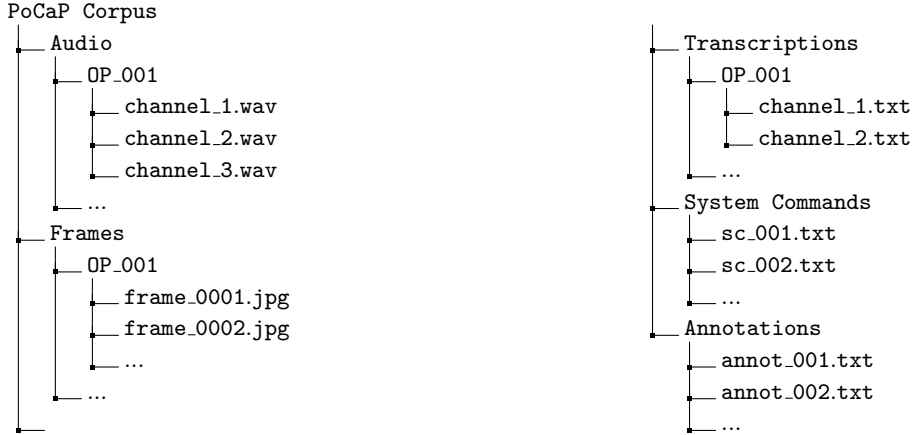

\begin{multicols}{2}
\dirtree{%
.1 PoCaP Corpus.
.2 Audio.
.3 OP$\_$001.
.4 channel$\_$1$.$wav.
.4 channel$\_$2$.$wav.
.4 channel$\_$3$.$wav.
.3 $...$.
.2 Frames.
.3 OP$\_$001.
.4 frame$\_$0001$.$jpg.
.4 frame$\_$0002$.$jpg.
.4 $...$.
.3 $...$.
.2  .
}

\dirtree{%
.1 .
.2 Transcriptions.
.3 OP$\_$001.
.4 channel$\_$1$.$txt.
.4 channel$\_$2$.$txt.
.3 $...$.
.2 System Commands.
.3 sc$\_$001$.$txt.
.3 sc$\_$002$.$txt.
.3 $...$.
.2 Annotations.
.3 annot$\_$001$.$txt.
.3 annot$\_$002$.$txt.
.3 $...$.
}
\end{multicols}
\caption{The multimodal data structure of the PoCaP Corpus, consisting of speech,  transcriptions, X-ray images, system commands, and workflow annotations.}
\label{fig:data_struct}
\end{figure}

After the data collection process, we created a data structure with different modalities. We extracted two-channel audio signals from OBS captures and single-channel audio from ambient videos. We saved this audio data separately. As shown in Figure \ref{fig:image}, OBS recordings include software tool section on the right side. We cropped this section and converted videos to frames at $1 fps$. The cropped section did not include relevant information about the intervention. Finally, we included system commands in our corpus. We used extracted audio and video frames for the alignment of the data sources. Moreover, we used the audio signals to obtain transcriptions of the conversations as explained in Section \ref{sec:transcription}. The data structure of our corpus is shown in Figure \ref{fig:data_struct}.

\subsection{Alignment}
\label{sec:align}
In our data collection protocol, three different recording devices are used. X-ray images and audio signals are recorded with OBS on a  laptop, system commands are recorded on a local computer, and ambient videos are recorded with a single camera. Each recording is started from the relative source manually in the following order: ambient video recording, X-ray and audio capture with OBS, and system commands recording with ACSOS. Thus, it is necessary to align these data sources before continuing with the preprocessing and data annotation steps. 

We used the audio channels to align the ambient videos and OBS video captures. First, we randomly chose a reference channel from the personal microphones. Next, we computed the autocorrelation function of the reference channel and the third channel from the ambient videos. We used the position of the maximum correlation point to find the time difference between these sources. We padded the microphone signals at the beginning with zeros to align them with the third channel from the ambient videos. Additionally, due to limitations of the battery capacity of the ambient camera, we could not capture the last parts of every operation, if the intervention took longer than approximately one and a half hours. For these operations, we padded the audio signals with zeros at the end as well, in order to have the same length as the microphone signals.  

Finally, systems commands are aligned utilizing the OBS video captures. When the monitoring software ACSOS is started from the local computer, it is displayed on the same screen as the X-ray images. Therefore, the launch scene of ACSOS is also captured in the OBS videos. Lastly, since the OBS video capture started before the system command recording with ACSOS, we are also able to observe it's starting time on the OBS video captures.

\subsection{Transcription}
\label{sec:transcription}
After the alignment step, we used Voice Activity Detection (VAD) and ASR algorithms to convert speech signals to corresponding transcriptions.

Personal microphones are always placed closely to the talkers and have low background noise. In this case, the VAD task is close to a signal activity detection task. Therefore, we used simple energy and correlation-based thresholding to detect voice activities \cite{backstrom2017speech}. Initially, we applied \textit{30 ms} length Hann Window with \textit{50 \%} overlap. For each window, we calculated signal energy $\sigma ^2$ and autocorrelation $c_k$ at $lag = 1$ of microphone signal $x$ as: 

\begin{equation}
\sigma ^2 (x) = ||x||^2 = \Sigma_{k=0}^{N-1}x_k^2
\end{equation}

and 

\begin{equation}
c_k = \frac{E\{ x_k x_{k-1} \}}{E\{ x_k^2 \}},
\end{equation}

where $k \ \epsilon \ \{ 0,...,N-1 \}$ is the time index and $E$ is the expectation operator. Afterwards, we applied a threshold to detect voice activities.  

After the voice detection, we used publicly available, pre-trained German ASR models to transcribe the speech. To find the best performing ASR model for our case, we manually transcribed a total of six minutes of speech data from different operations with different surgeons and tested six different models. We compared models with our transcriptions using the basic Word Error Rate (WER) as a metric. After the evaluation of the results, we chose \textit{stt\_de\_conformer\_transducer\_large} model for the preprocessing pipeline. All results are shown in Table \ref{table:asr}.

\begin{table}
\caption{Performance of different ASR models on a short clip from our dataset.}\label{table:asr}
\centering
\begin{tabular}{|l|c|}
\hline
ASR Model & PoCaP Corpus WER (\%)\\
\hline
\textit{asr-crdnn-commonvoice-de} \cite{speechbrain} & 85.80\\
\textit{stt\_de\_quartznet15x5} \cite{kuchaiev2019nemo} & 56.46\\
\textit{stt\_de\_citrinet\_1024} \cite{kuchaiev2019nemo} & 29.87\\
\textit{stt\_de\_contextnet\_1024} \cite{kuchaiev2019nemo} & 26.96\\
\textit{stt\_de\_conformer\_ctc\_large} \cite{kuchaiev2019nemo} & 29.37\\
\textit{stt\_de\_conformer\_transducer\_large} \cite{kuchaiev2019nemo} & 11.52 \\
\hline
\end{tabular}
\end{table}

\section{Discussion}

In this section, we discuss the challenges in our data collection process and considerations in terms of future machine learning approaches with our corpus. In Section 3, the average duration and standard deviation of PoCaP intervention in our dataset were reported. The execution of an intervention can be affected naturally by many different variables. Some actions can be repeated several times or may require a longer time period due to complications. The expertise of surgeons is another noteworthy factor. The duration of an intervention is significantly affected by the experience level of the surgeon.

Since speech and audio data are the main components of the PoCaP Corpus, the performance of an ASR algorithm is crucial for any application. This performance is significantly affected by the use of medical words and dialect. Even with correct transcriptions, speech dialects can pose additional challenges for the following steps. It may be necessary to consider translating these transcriptions to standard language before employing a language model or extracting word vectors. It would be challenging to train such a translation model, similar to difficulties in low-resource language processing.

Moreover, for the development of a smart operating room speech assistant, it is necessary to classify surgery-related conversations and understand synonyms in the speech. Some conversations recorded with personal microphones may not be related to the intervention. These conversations can occur between the surgeons, the assistant, or other medical personnel in the OR. These conversations should be classified and removed as they carry information not related to the procedure and could rather be seen as noisy observations at the semantic level. After this removal, synonyms should be processed next. Here, different surgeons or the same surgeon at different time points, may use synonyms to refer to the same underlying meaning. These commands should be recognized accordingly as having the same meaning.

As depicted in Table \ref{table:phases}, X-Ray images are used in three phases and are typically utilized for a short duration in order to emit a minimal radiation dose to the patient. 
This also highlights that X-ray images are only taken a few times during a single operation especially considering the overall duration of the intervention. This fact makes them a very valuable source of information for the task of phase recognition of an intervention since they can be used as distinct landmarks for the identification of specific phases. However, this aspect should be further investigated in future works.

In order to develop robust machine learning algorithms, a large dataset with a large variety of variables is necessary. In our case, the number of operations, the number of medical institutions, and the number of surgeons are important. Currently, our corpus includes 31 interventions performed by six surgeons in a single medical center. For these reasons, we plan to expand our corpus with recordings of more interventions in the future.

\section{Conclusion}
In this paper, we presented the PoCaP Corpus, which consists of speech and audio signals, X-ray images, and system commands collected from 31 PoCaP interventions in the Radiology Department of University Hospital Erlangen, Germany. With this unique dataset, we aim to contribute to the development of a smart speech assistant in the operating room of the future. When compared to previous corpora for workflow analysis, the PoCaP Corpus differs in data types and multi-modality. To best of our knowledge, the PoCaP Corpus is the first dataset that includes speech and audio signals and X-ray images for the development of a Smart Speech Assistant. 

We described the data collection procedure, used systems, and preprocessing stages. Furthermore, we described the alignment process of the three different sources using audio signals and screen recordings. Later, we extracted X-ray images from screen recordings and transcriptions from speech signals, recorded by two personal microphones equipped to medical personnel in the OR and reported the ASR results. Finally, we reported challenges and possible future directions. Speech recognition is a more challenging task due to the terminology in a medical environment and speech dialects when compared to the standard ASR tasks. Therefore, the performance of ASR algorithms is worse than in the usual settings. These challenges may also manifest in the feature extraction tasks of future work. In addition to ASR performance improvements, correctly recognized dialect words should be translated into the standard German language. Synonymous medical terms will be also included in the training of feature extraction algorithms.

In the future, we plan to expand our dataset with more recordings. In the development of a Smart Speech Assistant, the PoCaP Corpus can be used for workflow analysis, recognition of commands given by surgeons, optimization of operation parameters, remaining surgery time estimation, or similar tasks. In that sense, the PoCaP Corpus enables the development of a wide variety of applications and thereby has the potential of making significant contributions to surgical data science.

\section{Acknowledgement}

We greatfully acknowledge funding for this study by Friedrich-Alexander-University Erlangen-Nuremberg, Medical Valley e.V. and Siemens Healthineers AG within the framework of d.hip campus. 

\bibliographystyle{splncs04}
\bibliography{ref}

\end{document}